\documentclass[aps,prd,reprint]{revtex4-1}		

\usepackage{graphicx}

\newcommand{\GeVc}{GeV/$c$}

\newcommand{\rarr}{\rightarrow}
\newcommand{\beq}{\begin{equation}}
\newcommand{\eeq}{\end{equation}}

\newcommand{\subs}[1]{$_{#1}$}
\newcommand{\sups}[1]{$^{#1}$}
\newcommand{\mr}[1]{\mathrm{#1}}

\usepackage[pdftex]{hyperref}

\hypersetup{
  pdftitle={Polarised target for Drell--Yan experiment in COMPASS at CERN, part I},%
  pdfauthor={J. Matousek},%
  pdfsubject={},%
  pdfkeywords={COMPASS, fixed-target, polarized},%
  pdfstartview={},%
  bookmarksopen=true, breaklinks=true, debug=true, %
  colorlinks=true, linkcolor=red, citecolor=blue, urlcolor=blue
}
\begin{document}
\title{Polarised target for Drell--Yan experiment in COMPASS at CERN, part I}

\author{J.~Matou\v{s}ek}
\thanks{Speaker}
\email{jan.matousek@cern.ch}
\affiliation{Charles University, Prague, Czech Republic}

\author{A.~Berlin}
\affiliation{Ruhr-University Bochum, Germany}

\author{N.~Doshita}
\affiliation{Yamagata University, Japan}

\author{M.~Finger}
\affiliation{Charles University, Prague, Czech Republic}

\author{M.~Finger, Jr.}
\affiliation{Charles University, Prague, Czech Republic}

\author{F.~Gautheron}
\affiliation{Ruhr-University Bochum, Germany}

\author{N.~Horikawa}
\affiliation{Chubu University, Japan}

\author{S.~Ishimoto}
\affiliation{KEK, Japan}

\author{T.~Iwata}
\affiliation{Yamagata University, Japan}

\author{Y.~Kisselev}
\affiliation{JINR, Dubna, Russia}

\author{J.~Koivuniemi}
\affiliation{Ruhr-University Bochum, Germany}
\affiliation{University of Illinois at Urbana-Champaign, USA}

\author{K.~Kondo}
\affiliation{Yamagata University, Japan}

\author{W.~Meyer}
\affiliation{Ruhr-University Bochum, Germany}

\author{Y.~Miyachi}
\affiliation{Yamagata University, Japan}

\author{H.~Matsuda}
\affiliation{Yamagata University, Japan}

\author{G.~Nukazuka}
\affiliation{Yamagata University, Japan}

\author{M.~Pe\v{s}ek}
\affiliation{Charles University, Prague, Czech Republic}

\author{G.~Reicherz}
\affiliation{Ruhr-University Bochum, Germany}

\author{H.~Suzuki}
\affiliation{Chubu University, Japan}

\author{T.~Tatsuro}
\affiliation{University of Miyazaki, Japan}

\date{\today}
\begin{abstract}
In the polarised Drell--Yan experiment at the COMPASS facility in CERN pion beam with momentum of 190~GeV/$c$ and intensity about 10\sups{8} pions/s interacted with transversely polarised NH\subs{3} target. Muon pairs produced in Drell--Yan process were detected. The measurement was done in 2015 as the 1st ever polarised Drell--Yan fixed target experiment. The hydrogen nuclei in the solid-state NH\subs{3} were polarised by dynamic nuclear polarisation in 2.5~T field of large-acceptance superconducting magnet. Large helium dilution cryostat was used to cool the target down below 100~mK. Polarisation of hydrogen nuclei reached during the data taking was about 80\%. Two oppositely polarised target cells, each 55~cm long and 4~cm in diameter were used.

Overview of COMPASS facility and the polarised target with emphasis on the dilution cryostat and magnet is given. Results of the polarisation measurement in the Drell--Yan run and overviews of the target material, cell and dynamic nuclear polarisation system are given in the part II.
\end{abstract}
\maketitle
\section{Introduction}
\label{sec:intro}

COMPASS is a collaboration operating fixed forward-scattering target experiments for various high energy physics channels. The facility is located in North area of CERN. The physics data-taking started in 2002\cite{compass:prop1} and in 2012 it entered its second phase expanding further the original scientific goals\cite{compass:prop2}.The main points of interest are nucleon structure studies, hadron spectroscopy and studies of chiral dynamics.

The facility uses secondary beam of hadrons (mainly $\pi^-$ or p) with energy up to 280~\GeVc, or tertiary $\mu^+$ or $\mu^-$ beam with energy up to 190~\GeVc. Arrangement of the target region differs for the various programmes. Large solid-state polarised target described in this text and elsewhere\cite{compass:2007:nima,koivuniemi:2015} was used for semi-inclusive deep-inelastic scattering (SIDIS)\cite{bacchetta:2007} and Drell--Yan\cite{arnold:2008} measurements. Long liquid-hydrogen target together with recoil proton detector installed around it or solid state nuclear targets were used as well\cite{compass:2015:nima}. To ensure wide momentum and angular acceptance and particle identification, COMPASS spectrometer is divided in two stages each having its own dipole magnet, set of tracking detectors, electromagnetic and hadronic calorimeter and a muon filter. In addition, there is Ring Imaging Cherenkov counter for hadron identification in the first large-angle stage.

\section{Drell--Yan programme}
\label{sec:DY}

One of the programmes in the phase~I was SIDIS of muons on transversely-polarised deuterons and protons (inside \sups{6}LiD and NH\subs{3} targets respectively) 
	$\mu^+ + N^\uparrow \rarr \mu^+ + h + X$.
Azimuthal spin-dependent asymmetries were studied; in this text we mention the Sivers asymmetry only. While it was found to be compatible with zero for pions and kaons produced on polarised deuterons\cite{compass:2009:td}, sizeable positive Sivers asymmetry was observed for positive pions and kaons produced on polarised protons\cite{compass:2014:tp}. The effect can be explained assuming correlation between transverse momenta of quarks inside the nucleon and transverse spin of the nucleon, described in terms of Sivers function\cite{sivers:1990}.

In the phase~II, COMPASS intends to complement the SIDIS measurements by studying the Drell--Yan reaction with 190~\GeVc\ pion beam and transversely polarised proton target
	$\pi^- + \mr{p}^\uparrow \rarr \mu^+ + \mu^- + X$.
The Sivers effect is expected to play role also in this reaction, but with the opposite sign of the Sivers function~\cite{collins:2002}; Verification of this prediction is one of the goals of the COMPASS Drell--Yan programme. In general, studies of the Drell--Yan reaction are essential for our understanding of universality (or rather well-defined non-universality) of the parton distribution functions.

Drell--Yan experiments are challenging because of relatively small cross-section. Therefore, COMPASS used hadron beam with intensity of about $10^8~\pi$/s, the highest allowed by radiation protection rules and other hardware restrictions. An alumina hadron absorber with tungsten beam plug was placed downstream of the polarised target to act as muon filter and lower detector occupancies and radiation levels. Of course the polarised target faces the full intensity of the beam. FLUKA\footnote{FLUKA, \url{http://www.fluka.org/}} simulation had shown that the target material should get an estimated dose of about 20~kGy in 220 days of running\cite{koivuniemi:2015}. NH\subs{3} is radiation-hard material\cite{goertz:2002}, but the intense beam introduces additional heat, which could accelerate relaxation of the polarization, and the radiation could also in principle damage other parts of the target system.

\section{Polarised target}
\label{sec:PT}

Large solid-state target provides the high rate of interactions needed for measurements of asymmetries in muon SIDIS and Drell--Yan reactions. The NH\subs{3} was doped by electron beam irradiation with typically $10^{-3}$--$10^{-4}$ free radicals per nucleus\cite{koivuniemi:2015}. The hydrogen nuclei can reach polarisation $P=0.8$--0.9 with dynamic nuclear polarisation (DNP) method and their fraction in the NH\subs{3} or dilution factor $f=0.176$. Therefore, it has high figure of merit $\propto P^2 f^2$, which makes it an optimal choice for COMPASS-like facilities\cite{goertz:2002}. The radiation hardness of ammonia is also important for the Drell--Yan experiment.

\begin{figure}
	\includegraphics[width=0.48\textwidth]{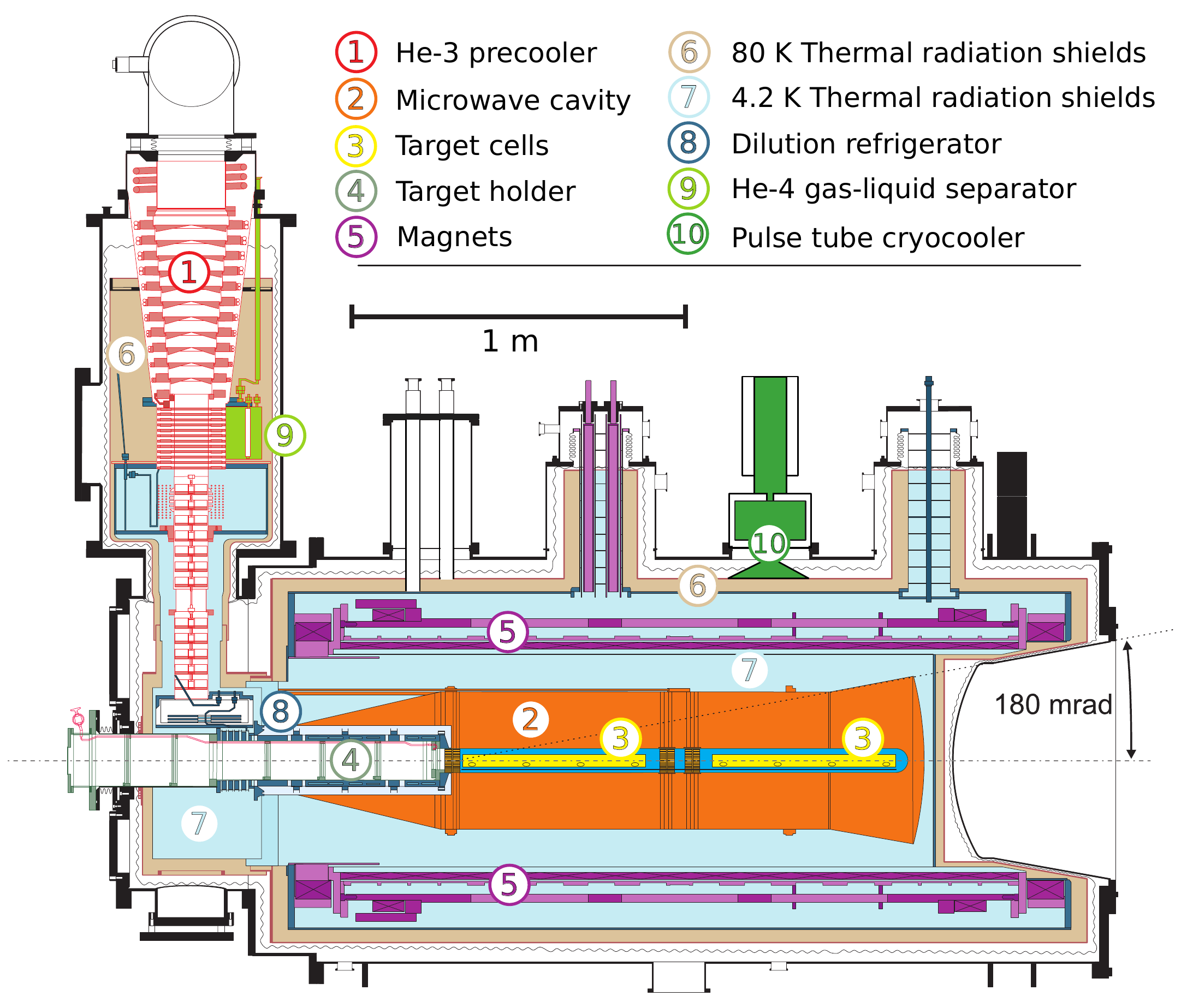}
	\caption{Polarised target cryostat with key elements.}
	\label{fig:PT}
\end{figure}

The DNP method requires high cooling power at temperature lower than 1~K and high and homogeneous magnetic field. Such magnetic field would be difficult to achieve in transverse polarisation mode. Therefore, frozen spin technique is used. After the target material reaches desired polarisation in longitudinal magnetic field, the DNP is switched off and the target material is rapidly cooled below 100~mK. At this temperature the relaxation time of the polarisation reaches order of 1\,000 hours. Then the magnetic field can be rotated to the transverse direction. \sups{3}He--\sups{4}He dilution cryostat (Sec.~\ref{sec:cryostat}) can meet the cooling needs in both DNP and frozen spin regimes. Large-aperture superconducting magnet (Sec.~\ref{sec:magnet}) with solenoid and dipole coils provides the magnetic field. The target cryostat with the key parts is shown on Fig.~\ref{fig:PT}. Polarisation measurement is done by continuous-wave nuclear magnetic resonance (NMR) system, described in the part II of this text\cite{nukazuka:2016} together with the DNP system, target cell construction and polarisation measured in the Drell--Yan run in 2015.

Many parts of the COMPASS polarised target come from the target of the Spin muon collaboration (SMC), which started operation in 1992\cite{smc:1999}. Around 2000 it was modified for COMPASS\cite{compass:2007:nima} and in 2006 got new large-acceptance magnet\cite{gautheron:2007}. In 2014 the magnet was refurbished, monitoring system of the cryostat was upgraded and new target cells for the Drell--Yan experiment were produced.

\section{Dilution cryostat}
\label{sec:cryostat}

Large mixing chamber about 1.6~m long and 7~cm in diameter is needed to contain the target cells. About 9\,000~l of He gas mixture with 10--15\% of \sups{3}He is used for operation of the refrigerator. The \sups{3}He-rich vapour is pumped from the still by 8 Pfeiffer roots blowers in series. The original copper heat exchangers of the pumps, damaged by ageing, were replaced by new ones made of stainless steel\cite{koivuniemi:2015}. The compressed gas goes through activated-charcoal filters at room and liquid nitrogen temperatures to remove impurities. The gas returns to the mixing chamber through pre-cooler and series of flattened stainless steel tubes (0.1~m\sups{2} continuous heat exchanger) and sintered copper heat exchangers (12~m\sups{2} step heat exchanger).

\begin{figure}
	\includegraphics[width=0.48\textwidth]{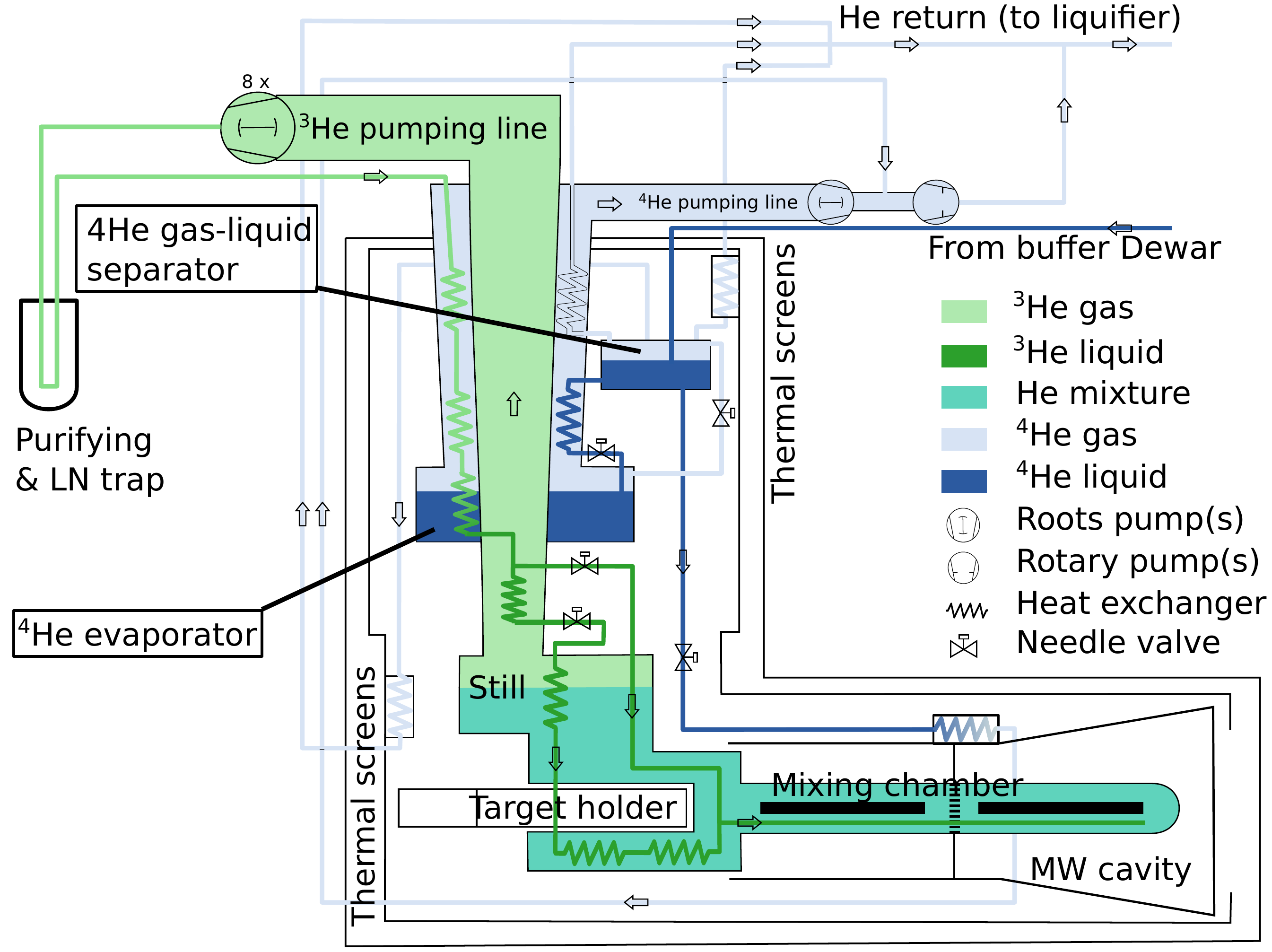}
	\caption{Dilution refrigerator -- scheme of operation.}
	\label{fig:DR}
\end{figure}

The \sups{3}He flow rate and thus the cooling power is controlled with an electric heater placed on the bottom of the still. In the frozen-spin mode the still temperature is about 1.0~K and pressure 0.42~mbar with flow of 0.07~g/s and 0.2~mbar roots inlet pressure\cite{koivuniemi:2015}. Temperature below 50~mK was measured in the mixing chamber without beam and with beam of about $10^{8}~\mu$/s, which represents estimated heat input of 1~mW\cite{doshita:2004}. In 2015 with the hadron beam of similar intensity the temperature was about 30--40~mK higher\cite{koivuniemi:2015}. In the DNP mode the cooling power is about 350~mW at 0.3~K with about 3~times higher \sups{3}He flow rate\cite{doshita:2004}.

In addition to the circulating mixture the cryostat consumes about 15--20~l/h of liquid \sups{4}He for the precooler and cooling of thermal screens\cite{doshita:2004}. The precooler includes \sups{4}He evaporator at 1.4~K pumped by Leybold roots blower and Sogevac rotary pump with flow of 0.4 g/s\cite{koivuniemi:2015}. Diagram of the dilution refrigerator is shown on Fig.~\ref{fig:DR}.

For monitoring of the operation the cryostat is instrumented by a set of thermometers. Temperatures above 4~K are measured by diode thermometers read by Lakeshore LS218 temperature monitor. The dilution chamber is equipped by 3 RuO and 3 Speer resistive thermometers read by Picowatt AVS-46 and AVS-47 resistance bridges. These instruments together with vacuum gauges are periodically read by Perl scripts running on Linux computer together with custom DIM\footnote{Distributed information management system, \url{http://dim.web.cern.ch/}} server written in C++, which publishes the measured values over local network to make them available for the centralised COMPASS Detector control system. In addition, the data can be stored in MySQl or SQLite databases. This software was developed for the 2015 run to improve remote monitoring.

\section{Magnet}
\label{sec:magnet}

The present magnet was put in operation in 2006\cite{gautheron:2007}. It is 2\,350~mm long and its large aperture with 638~mm in diameter is essential for the acceptance of the COMPASS spectrometer. The magnet contains superconducting solenoid coil complemented by 16 superconducting shim coils that produces 2.5~T field parallel to the beam axis with homogeneity better than $10^{-5}$ over the volume of the target cells. The longitudinal field is used for the DNP. There is also superconducting dipole saddle coil capable of producing 0.63~T transverse field, which is used in the frozen-spin mode to rotate the polarisation to the transverse direction and then to hold it. The polarisation can be rotated this way by $180^\circ$ (takes about 1~h) without the need of DNP reversal (takes at least 24~h) -- a great advantage for runs with longitudinally polarised target. For the transverse runs it is not possible because the beamline would have to be re-adjusted in order to compensate the influence of the reversed dipole field. The solenoid and dipole power converters are made of two Delta Electronica one quadrant
15 V/400 A power supplies connected in parallel, followed by polarity switch.

Between 2011 and 2015 the magnet underwent general refurbishment by CERN magnet team. Shorts on two shim coils were repaired and several design weaknesses were addressed to improve safety and lower service costs. A CryoMech pulsed tube cryocooler PT60-UL with CP830 air cooled compressor package was installed into a new turret (Fig.~\ref{fig:PT}). It is used for cooling of the magnet thermal screens to 60~K, which reduces the liquid He consumption to about 10~l/h\cite{koivuniemi:2015}.

The magnet was instrumented by 9 Pt1000 temperature sensors and 9 standard CERN LHe temperature bridges. The solenoid is monitored by 4 and the dipole by 3 voltage bridges for quench detection. CERN UNICOS\footnote{Unified industrial control system, \url{http://unicos.web.cern.ch/}} is used for monitoring and control of the magnet. The system publishes the measured parameters by DIP\footnote{DIP, \url{http://cern.ch/dip}} server, which makes them available for the COMPASS Detector control system. If the quench detection circuit sees sudden voltage rise, the magnet safety system opens circuit breakers from the power converters and at the same time switches on 16 heaters attached to the magnet coils. External diode and resistive elements are used to dissipate part of the stored energy.

\section{Conclusion}
\label{sec:conclusion}

Study of the Drell--Yan interaction of pion beam and transversely polarised proton target may shed more light to our understanding of the parton distribution functions and it is a corner stone of the COMPASS~II physics programme\cite{compass:prop2}. The measurement was done in 2015 as the first-ever polarised Drell--Yan experiment and the first results are expected soon\cite{quaresma:2016}. The polarised target was an essential part of the experiment. Both the dilution cryostat and the superconducting magnet of the target were refurbished before the 2015 run. The temperature in the mixing chamber exposed to the beam of $10^{8}~\pi$/s was bout 30~mK higher than with $\mu$ beam of similar intensity. No radiation damage was seen in the target system.

Construction of the new target cells, DNP system and measurement of the polarization in the 2015 run are discussed in the part II of this text\cite{nukazuka:2016}.

\bibliography{matousek}{}
\end{document}